\documentclass[12pt]{article}
\usepackage{graphicx}
\usepackage{epsf,amsmath,bbold,amsfonts,stmaryrd}

\usepackage[utf8]{inputenc}
\usepackage{mathrsfs}
\usepackage{appendix}
\usepackage{amssymb}
\usepackage{float}
\usepackage{color} 
\usepackage{cite}
\usepackage[colorlinks]{hyperref}
\hypersetup{pageanchor=false}
\usepackage{indentfirst}
\usepackage{url}
\usepackage{float}
\usepackage{caption}
\usepackage[numbers,square,comma,sort&compress,merge]{natbib}

\input epsf
\allowdisplaybreaks

\newcommand{\be}{\begin{equation}}
\newcommand{\ee}{\end{equation}}

\renewcommand{\a}{\alpha}
\renewcommand{\b}{\beta}
\newcommand{\g}{\gamma}
\newcommand{\G}{\Gamma}
\renewcommand{\d}{\delta}

\newcommand{\z}{\zeta}

\renewcommand{\k}{\kappa}
\renewcommand{\l}{\lambda}
\newcommand{\m}{\mu}
\newcommand{\n}{\nu}
\renewcommand{\j}{\xi}

\newcommand{\p}{\pi}
\renewcommand{\r}{\rho}
\newcommand{\s}{\sigma}

\newcommand{\f}{\phi}
\newcommand{\x}{\chi}

\begin{document}

\begin{titlepage}
\vspace{5cm}

\vspace{2cm}

\begin{center}
\bf \Large{Scale invariant alternatives to general relativity. II. 
Dilaton properties}

\end{center}

\begin{center}
{\textsc {Georgios K. Karananas, Mikhail Shaposhnikov}}
\end{center}

\begin{center}

{\it Laboratory of Particle Physics and Cosmology\\
 Institute of  Physics \\
\'Ecole Polytechnique F\'ed\'erale de Lausanne \\ 
CH-1015 Lausanne\\
 Switzerland}
\end{center}

\begin{center}
\texttt{\small georgios.karananas@epfl.ch} \\
\texttt{\small mikhail.shaposhnikov@epfl.ch} 
\end{center}

\vspace{2cm}

\begin{abstract}

In the present paper, we revisit gravitational theories which are invariant under TDiffs -- transverse (volume preserving) diffeomorphisms and global scale transformations. It is known that these theories can be rewritten in an equivalent diffeomorphism-invariant form with an action including an integration constant (cosmological constant for the particular case of non-scale-invariant unimodular gravity). The presence of this integration constant, in general, breaks explicitly scale invariance and induces a runaway potential for the (otherwise massless) dilaton, associated with the determinant of the metric tensor.  We show, however,  that if the metric carries mass dimension $\left[\text{GeV}\right]^{-2}$, the scale invariance of the system is preserved, unlike the situation in theories in which the metric has mass dimension different from $-2$. The dilaton remains massless and couples to other fields only through derivatives, without any conflict with observations. We observe that one can define a specific limit for fields and their derivatives (in particular, the dilaton goes to zero, potentially related to the small distance domain of the theory) in which the only singular terms in the action correspond to the Higgs mass and the cosmological constant. We speculate that the self-consistency of the theory may require the regularity of the action, leading to the absence of the bare Higgs mass and cosmological  constant, whereas their small finite values may be generated by nonperturbative effects.
 
\end{abstract}

\end{titlepage}

\section{Introduction}
\label{sec:Intro}

It is well known that a self-consistent gravitational theory does not require invariance under the full group of diffeomorphisms~\cite{Buchmuller:1988wx,Alvarez:2006uu}. Rather, it is enough to consider the subgroup of the coordinate transformations with Jacobian equal to unity
\be
\label{tdiffs-def}
x'=F(x)\ ,~~~\text{such that}~~~J\equiv \left|\frac{\partial F}{\partial x}\right|=1\ ,
\ee
which constitute the \emph{transverse diffeomorphisms} (TDiffs), also called volume preserving diffeomorphisms. As one might expect, theories invariant under TDiffs contain -- in addition to the two polarizations of the massless graviton -- an extra propagating scalar mode associated with the determinant of the metric.\footnote{It is possible to eliminate this extra degree of freedom by forcing the determinant to take a constant value, like for example in Unimodular Gravity where it is fixed to be equal to one~\cite{vanderBij:1981ym,Unruh:1988in,Henneaux:1989zc}.} This minimalistic approach to gravitational dynamics, once combined with the requirement of exact  scale invariance, results into an interesting class of theories (for which in what follows we will use the acronym SITDiff) that  were constructed and studied in detail in~\cite{Blas:2011ac}. 

The results of ~\cite{Blas:2011ac}  can be summarized as follows. In this class of theories, the scalar degree of freedom related to the metric determinant is identified with a massless dilaton $\s$ that only couples derivatively and thus evades the fifth force constraints. Assuming that the metric is dimensionless and the Lagrangian contains up to two derivatives of the fields, the most general scalar-tensor theory that includes matter fields was presented. The form of the action can not be completely fixed;  rather, it involves arbitrary functions of the metric determinant (``theory defining functions''),  since this quantity behaves as a scalar under the restricted coordinate transformations. It was shown that the invariance of the system under dilatations
\be
\label{scal-trans-fields-norm}
x^\m\rightarrow \a^{-1} x^\m~~~\text{and}~~~ g_{\m\n}(x)\rightarrow  g_{\m\n}(\a^{-1} x) \ ,
\ee
with $\a$ a constant, is explicitly broken at the level of the equations of motion by an arbitrary integration constant that appears because of TDiff rather than Diff invariance. This gives rise to a run-away potential for the dilaton. It was demonstrated that by appropriately choosing the theory defining functions, it is possible to get a theory which has interesting implications for particle physics and cosmology.  Its particle physics sector can be made identical to the Standard Model, whereas it is able to account for the inflationary period in the early Universe and provide a natural candidate for dynamical dark energy.

In this paper we generalize the aforementioned work by investigating what are the implications on the structure of these models when the metric tensor $g_{\m\n}$ has (arbitrary) mass dimension. Usually, it is somehow taken for granted that $g_{\m\n}$  is dimensionless, whereas the coordinates $x^\m$ carry dimensions of length. However, this is nothing more than a particular choice which follows ``naturally'' only when the Minkowski space-time is described in terms of cartesian coordinates. Notice that this choice is certainly not the most appropriate one when other coordinate systems are used, let alone when curved space-times are considered. 

Let us carry out some elementary dimensional analysis. Although what follows is in a sense trivial if the theory under consideration is diffeomorphism invariant, the situation changes considerably for SITDiff theories, since the metric determinant is a propagating degree of freedom that plays the role of the dilaton. By definition, $\left[g_{\m\n}dx^\m dx^\n\right]=\left[\text{GeV}\right]^{-2}$, so in principle, we have  the liberty to assign arbitrary dimensions -- also fractional -- both to $x^\m$ and $g_{\m\n}$, i.e. 
\be
\label{assign-dims}
\left[ x^\m\right]=\left[\text{GeV}\right]^{-p}\ ,~~~\left[g_{\m\n}\right]=\left[\text{GeV}\right]^{-2q} \ ,
\ee
as long as $p+q=1$. The dilatations now act on the coordinates and the metric as 
\be
\label{scal-trans-fields}
x^\m\rightarrow \a^{-p} x^\m~~~\text{and}~~~ g_{\m\n}(x)\rightarrow \a^{-2q} g_{\m\n}(\a^{-p} x) \ ,
\ee
since the scaling dimensions coincide with the mass dimensions.  Of a special interest  is the case in which $x^\m$ merely label events on the manifold and the metric carries dimensions of area
\be
\label{p0q1}
p=0~~\text{and}~~~q=1\ .
\ee

We will see that the class of  theories with $p\neq 0$ is equivalent to that already described in ~\cite{Blas:2011ac}. However, the case~\eqref{p0q1} is different. In particular,  a dilatation symmetry breaking potential for the dilaton will be shown to be  absent, an otherwise generic feature of the  theories with $p\neq 0$. Moreover, it is remarkable that by abandoning the prejudice of a dimensionless metric and requiring that there are no terms with more than two derivatives in the action, we can completely fix its form for pure gravity without matter fields.  It should be noted that, in principle, one can relax the requirement of having an action that contains terms which are at most quadratic in the derivatives. To ensure absence of ghosts, the starting point in this case should either be Horndeski theory~\cite{Horndeski:1974wa} or $f(R)$-gravity, see~\cite{Sotiriou:2008rp} and references therein.\footnote{The Horndeski theory is the most general scalar-tensor action with second order equations of motion. The scale- and Weyl-invariant subclasses of this theory have been identified in~\cite{Padilla:2013jza}. It would be interesting to understand what are the implications of having invariance under TDiff instead of the full group of diffeomorphisms, an investigation we leave for elsewhere.} For the latter, we will show that they can be used as the starting point for constructing biscalar SITDiff theories.

Next, we present how a scalar field can be incorporated in a consistent manner. If this field is identified with the Standard Model Higgs boson, we end up with a phenomenologically viable SITDiff theory. As we will demonstrate, the Higgs mass as well as the cosmological constant appear in the action in a peculiar way, different from the other terms.

Inspired by this, we formulate a set of rules that allows us to distinguish formally the Higgs mass and the cosmological constant from other contributions to the action based on their behaviour when the dilaton goes to zero. Since this field is related to the metric determinant that now carries dimension of length, this limit potentially corresponds to vanishing length and thus it is in a sense related to the UV regime.  More precisely, we notice that when the theory is expressed in terms of variables that are conjugate to the time and space derivatives of the fields (canonical four-momenta \cite{Schwinger:1951xk,Schwinger:1953tb}), then the only terms which  involve inverse powers of the dilaton --  and thus are presumably singular at the UV limit -- are the Higgs boson mass and the cosmological constant. Based on that, we speculate that their absence in the action may be a requirement of the self-consistency of the theory in the UV domain. The smallness of the observed  low energy  values of the Higgs mass and of the cosmological constant,  perhaps,  could be attributed to  some yet unknown nonperturbative mechanism. 

This paper is structured as follows. In Sec.~\ref{sec:dil}, we construct the most general SITDiff theory that contains only the dilaton and  study its properties. In Sec.~\ref{sec:hig-dil}, we demonstrate how matter fields are introduced in this framework. We present a phenomenologically viable model that in addition to the dilaton contains an extra scalar field, that is identified with the Standard Model Higgs boson. In Sec.~\ref{sec:Regul}, we formulate the assumptions that make it possible to single out  the presence of certain terms in the action by requiring that the theory has a regular limit when determinant of the metric goes to zero. We present our conclusions in Sec.~\ref{sec:conclus}. In Appendix~\ref{sec:dim-an}, we present for completeness  the dimensionality of various geometrical objects.

\section{Pure gravity}
\label{sec:dil}

As a warm-up exercise, we will write down the most general theory that contains at most two derivatives of the fields and is invariant under the restricted coordinate transformations and dilatations, which are given, respectively, by~\eqref{tdiffs-def} and~\eqref{scal-trans-fields}. The unique action that satisfies these requirements reads
\be
\label{TDSI-1}
S=\int d^4 x\sqrt{g}\left[ \frac{\z}{2} g^{\frac{1}{4(p-1)}} R - c_1\, g^{\frac{1}{4(p-1)}-2}g^{\m\n}\partial_\m g \partial_\n g -c_2 g^\frac{1}{2(p-1)} \right] \ ,
\ee
where  $\z, \ c_1,\ \text{and} \ c_2$ are dimensionless constants and the scalar curvature $R$ is defined in Appendix~\ref{sec:dim-an}. Observe that for  $p=1$, the above expression becomes singular. This is a manifestation of the fact that if we consider the standard mass (and scaling) dimension for the metric and coordinates, it is not possible to construct SITDiff theories with the metric determinant only. This was also realized in~\cite{Blas:2011ac}.

To get a better grasp on the dynamics of this theory, it is desirable to recast it in a form invariant under the full group of diffeomorphisms. Once we consider a coordinate transformation with $J\neq 1$, we obtain
\be
\label{TDSI-2}
S=\int d^4 x \sqrt{g}\left[\frac{\z}{2} \sigma^{\frac{1}{4(p-1)}} R - c_1\, \sigma^{\frac{1}{4(p-1)}-2}g^{\m\n}\partial_\m \sigma \partial_\n \sigma -c_2 \sigma^\frac{1}{2(p-1)}+c_3\,\sigma^{-1/2}\right] \ ,
\ee
where we defined the dilaton field $\sigma\equiv J^2g$, a scalar under diffeomorphisms. 
Some comments are in order at this point. First of all, when the theory is written this way, its particle spectrum can be read off immediately. It contains, in total, three degrees of freedom: the two graviton polarizations and an additional  scalar field which is associated with the determinant of the metric. Moreover, we notice the appearance of an extra term in the action proportional to the integration constant $c_3$, which emerged through the equations of motion; see for example~\cite{Alvarez:2006uu,Shaposhnikov:2008xb,Blas:2011ac} and references therein. It should be noted that for $p\neq 0$ (and equivalently $q\neq 1$), the resulting theories are all equivalent to the ones which were already considered in~\cite{Blas:2011ac}. In this case, the aforementioned constant necessarily carries  dimensions and consequently, its presence explicitly breaks the symmetry of the theory under dilatations and produces a run-away potential for the dilaton. This is a generic feature of these models. Hence, it seems that $p=0$ is a rather special point in the phase space of the theory, since $c_3$ is dimensionless and the theory under consideration is exactly scale invariant.\footnote{Actually, it  coincides with the induced gravity model introduced in~\cite{Zee:1978wi,Smolin:1979uz}. } 

Let us now introduce a field $\x$ with canonical dimensions,
\be
\label{sigm-chi-1}
\x=\s^{\frac{1}{8(p-1)}} \ , 
\ee
and set 
\be
\label{c1}
c_1=\frac{1}{128(p-1)^2} \ ,
\ee
so that~\eqref{TDSI-2} is equivalently rewritten as 
\be
\label{TDSI-3}
S=\int d^4 x \sqrt{g}\left[\frac{\z}{2}  \x^2 R -\frac{1}{2} g^{\m\n}\partial_\m \x \partial_\n \x -c_2\,\x^4+c_3\,\x^{-4(p-1)}\right] \ . 
\ee

In order to eliminate the mixing between the field and the curvature, it is convenient to write the theory  such that the gravitational part takes the standard Einstein-Hilbert form and all nonlinearities are moved to the scalar sector. To this end, we perform the following change of variables,
\be
\label{conf-trans}
g_{\m\n}\rightarrow \omega^{-2}g_{\m\n} \ ,~~~\text{with}~~~\omega=\frac{\sqrt{\z}\x}{M_P} \ ,
\ee
where $M_P=2.4\times 10^{18} \ \text{GeV}$ is the Planck mass. A straightforward calculation gives us the action in the Einstein frame:
\be
\label{ein-fr-1}
S=\int d^4x\sqrt{g}\left[\frac{M_P^2}{2}R-\frac{M_P^2}{2\z}\left(1+6\z\right)\x^{-2} g^{\m\n}\partial_\m \x \partial_\n \x -\frac{c_2 M_P^4}{\z^2}+\frac{c_3 M_P^4}{\z^2}\x^{-4p} \right] \ .
\ee
To bring the kinetic term for the field into canonical form,  we define
\be
\label{chi-phi}
\x=e^{\frac{\gamma \f}{M_P}} \ , \ \ \ \g=\sqrt{\frac{\z}{1+6\z}} \ ,
\ee
so that~\eqref{ein-fr-1} becomes
\be
\label{ein-fr-2}
S=\int d^4x\sqrt{g}\left[\frac{M_P^2}{2}R-\frac{1}{2}g^{\m\n}\partial_\m \f \partial_\n \f -\frac{c_2 M_P^4}{\z^2}+\frac{c_3 M_P^4}{\z^2}e^{\frac{-4p\gamma \f}{M_P}}\right] \ .
\ee

We observe that for $p=0$, the theory in the Einstein frame boils down to that of a massless minimally coupled scalar field in curved spacetime,
\be
\label{ein-fr-3}
S=\int d^4x\sqrt{g}\left[\frac{M_P^2}{2}R-\frac{1}{2}g^{\m\n}\partial_\m \f \partial_\n \f -\frac{c M_P^4}{\z^2}\right] \ ,
\ee
where we denoted $c=c_2-c_3$. Notice that the (exact) scale invariance of the model in the Jordan frame has manifested itself as an (exact) shift symmetry,
\be
\label{shift-sym}
\f\rightarrow \f +\text{constant} \ ,
\ee
when the theory was written in the Einstein frame. Thus, instead of the typical symmetry-breaking exponential potential for the field, we got a contribution to the cosmological constant term. This is a novel feature of SITDiff theories with dimensionless coordinates.

At this point it is worth taking a short detour and discussing the implications of allowing terms with more than two derivatives of the fields in the action, even though it lies outside the main scope of this paper. In general, higher-derivative terms may put the self-consistency of a theory under scrutiny, since their presence often (but not always) leads to the appearance of ghostly degrees of freedom in the spectrum. One of the simplest examples of healthy theories that involve an arbitrary number of derivatives of the metric in the action is ``$f(R)$ gravity''~\cite{Sotiriou:2008rp}. It is based on the replacement of the Einstein-Hilbert term which is linear in the scalar curvature, by an arbitrary function of $R$, such that the action reads
\be
\label{f_R_1}
S=\frac{M_P^4}{2}\int d^4 x \sqrt{g}f\left(R\right) \ ,
\ee
where $f(R)$ need not be local and for dimensional reasons can only depend on $R/M_P^2$. This modification to general relativity is motivated both from theory and phenomenology. Since gravity is an effective field theory, curvature corrections are expected to be present and play significant role when quantum effects are taken into account. Also, with an appropriate choice of the  function, it is possible to get interesting cosmological consequences for the early and late Universe.\footnote{The succesful Starobinsky model of inflation~\cite{Starobinsky:1980te} is a higher-derivative theory with 
$$
f\left(R\right)=\frac{R}{M_P^2}+ \frac{2\a R^2}{M_P^4} \ ,
$$
and $\alpha>0$ is a dimensionless constant.}

As is customary when dealing with these theories, it is convenient to express the above in a way that the dynamics of the extra degree(s) of freedom is separated from the gravitational sector. Performing a Legendre transformation, we can cast~\eqref{f_R_1} into the following equivalent form,
\be
\label{f_R_2}
S=\frac{M_P^4}{2}\int d^4 x \sqrt{g}\Big[f'(\chi) R-V(\chi)\Big] \ ,
\ee
where prime denotes derivative with respect to $\chi$ and we define
\be
\label{f_R_pot}
V(\chi)=\chi\, f'(\chi)-f(\chi) \ . 
\ee
Note that the absence of ghosts forces us to impose $f'(\chi)>0$, and we have to require $f''(\chi)\neq 0$ such that $\chi=R$. 

To make the kinetic term for $f'(\chi)$ appear explicitly in the action, we Weyl-rescale the metric as 
\be
\label{star_infl_5}
g_{\m\n}\rightarrow \frac{1}{M_P^2f'(\chi)}g_{\m\n} \ ,
\ee
to obtain
\be
\label{star_infl_6}
S=\int d^4 x \sqrt{g}\left[\frac{M_P^2}{2}R-\frac{3M_P^2}{4f^{'2}(\chi)}\partial_\m f'(\chi)\partial^\m f'(\chi)-\frac{V(\chi)}{2f^{'2}(\chi)} \right] \ . 
\ee
Finally, we introduce
\be
\label{star_infl_7}
\varphi=\sqrt{\frac{3}{2}}M_P \log\left[M_P^2f'(\chi)\right] \ ,
\ee
in terms of which the action takes its ``standard'' form,
\be
\label{star_infl_8}
S=\int d^4 x \sqrt{g}\left[\frac{M_P^2}{2}R-\frac{1}{2}(\partial_\m \varphi)^2-U(\varphi) \right] \ ,
\ee
with
\be
U(\varphi)=\frac{V[\chi(\varphi)]}{2f^{'2}[\chi(\varphi)]} \ .
\ee

The above procedure can be straightforwardly generalized to the class of theories that we are considering here, something that will lead to biscalar theories. For the purposes of illustration, it suffices to stick to the ``special'' case $p=0$. Requiring invariance under dilatations and TDiff fixes the action as
\be
\label{f_R_Tdiff}
S=\int d^4 x\Big[ f\left(R\right)- c_1\, g^{-7/4}g^{\m\n}\partial_\m g \partial_\n g -c_2 \Big]\ ,
\ee
where, for the function $f$ to be dimensionless, the scalar curvature must only appear multiplied by $g^{1/4}$. Repeating the steps outlined previously and  restoring the invariance under general coordinate transformations, we can write the above as
\be
S=\int d^4 x\sqrt{g}\Big[\s^{-1/4} f'\left(\chi\right)R-\s^{-1/4}V(\chi)- c_1\, \s^{-9/4}g^{\m\n}\partial_\m \s \partial_\n \s -c\, \s^{-1/2}\Big]\ ,
\ee
where, as before, $\s=J^2g$, $c=c_2-c_3$, and the ``potential'' $V(\chi)$ was presented in~\eqref{f_R_pot}. As expected, we ended up with a scalar-tensor theory that contains -- on top of the graviton -- two propagating fields. Choosing the function in~\eqref{f_R_Tdiff} appropriately, it is possible to construct a vast number of models with interesting cosmological phenomenology.

\section{Including matter fields}
\label{sec:hig-dil}

In the present section we wish to generalize the SITDiff theory we constructed previously by showing how matter fields can be incorporated into this setup. Let us start by introducing another scalar $h$ with canonical mass dimensions.

We saw that the theory presented previously was completely determined by requiring invariance under TDiff and scale transformations; see~\eqref{tdiffs-def} and~\eqref{scal-trans-fields}, respectively. When we bring into the game an extra scalar field, the situation changes. The dimensionless quantity,
\be
\label{dim-quant}
h^2g^{-\frac{1}{4(p-1)}}\ ,
\ee
is invariant under both TDiff and dilatations. Therefore, arbitrary functions of the above can, in principle, appear in the action. As in the previous section, we restrict ourselves to terms that are, at most, quadratic in the derivatives of the various fields. Dimensional analysis dictates that the gravitational and scalar sectors of the action that possess the desired properties read
\be
\begin{aligned}
\label{TD-SI2-full-1}
S=\int d^4x \sqrt{g} &\left[\frac{\z}{2}g^{\frac{1}{4(p-1)}}F_1\left(h^2g^{-\frac{1}{4(p-1)}}\right)R-c_1 g^{\frac{1}{4(p-1)}-2}F_2\left(h^2g^{-\frac{1}{4(p-1)}}\right)g^{\m\n}\partial_\m g \partial_\n g\right.\\
&\left.-\frac{1}{2} F_3\left(h^2g^{-\frac{1}{4(p-1)}}\right)g^{\m\n}\partial_\m h \partial_\n h+\d g^{-1}h F_4\left(h^2g^{-\frac{1}{4(p-1)}}\right)g^{\m\n}\partial_\m g \partial_\n h\right.\\
&\left.-c_2 g^{\frac{1}{2(p-1)}}V\left(h^2g^{-\frac{1}{4(p-1)}}\right)\right] \ .
\end{aligned}
\ee
Here $F_i$ and $V$ are arbitrary functions that can only depend on the dimensionless combination~\eqref{dim-quant}.  For later convenience, we have also included the constants $\z,c_1,c_2,$ and $\d$. We now consider a transformation with $J\neq 1$ and introduce $\sigma=J^2 g$ to recast the action into its diffeomorphism-invariant form:
\be
\begin{aligned}
\label{TD-SI2-full-2}
S=\int d^4x \sqrt{g} &\left[\frac{\z}{2}\s^{\frac{1}{4(p-1)}}F_1\left(h^2\s^{-\frac{1}{4(p-1)}}\right)R-c_1 \s^{\frac{1}{4(p-1)}-2}F_2\left(h^2\s^{-\frac{1}{4(p-1)}}\right)g^{\m\n}\partial_\m \s \partial_\n \s\right.\\
&\left.-\frac{1}{2} F_3\left(h^2\s^{-\frac{1}{4(p-1)}}\right)g^{\m\n}\partial_\m h \partial_\n h+\d\s^{-1}h F_4\left(h^2\s^{-\frac{1}{4(p-1)}}\right)g^{\m\n}\partial_\m \s \partial_\n h\right.\\
&\left.-c_2 \s^{\frac{1}{2(p-1)}}V\left(h^2\s^{-\frac{1}{4(p-1)}}\right)+c_3\,\sigma^{-1/2}\right] \ .
\end{aligned}
\ee
We should stress, once again, that unless $p=0$, the above theory is completely analogous to the one presented in~\cite{Blas:2011ac}, in which the term proportional to $c_3$ explicitly violates the invariance of the theory under scale transformations. Also, like in the purely gravitational theory, the limit $p=1$ is peculiar. In the two-field case, however, the presence of the extra scalar makes it possible to construct SITDiff theories even if the dimensionality of the metric is zero.

Before moving on, we would like to mention that the inclusion of gauge fields and fermions in the present framework goes along the same lines as in~\cite{Blas:2011ac}. Since here we are interested solely on the gravitational and scalar sectors of the SITDiff theories, the interested reader is referred to this work for an extensive discussion on the subject.

\subsection{Higgs-dilaton cosmology from TDiff} 

The presence of gravity in the theory under consideration makes it nonrenormalizable. Hence, it should be thought  of as an effective field theory which is valid up to some energy scale. Let us assume that for energies well below this cutoff, $h\ll \s^{\frac{1}{8(p-1)}}$. In this case, if the the various functions are analytic in their argument, we can Taylor expand them as
\be
\begin{aligned}
F_i(h^2\s^{-\frac{1}{4(p-1)}})&\approx 1+f_i\, h^2\s^{-\frac{1}{4(p-1)}}+\ldots \ ,\\
V(h^2\s^{-\frac{1}{4(p-1)}})&\approx 1+\a h^2\s^{-\frac{1}{4(p-1)}}+\b h^4\s^{-\frac{1}{2(p-1)}}+\ldots \ ,
\end{aligned}
\ee
where the ellipses denote higher order terms, and $f_i,\a,\b$ are constants that depend on the structure of the particular function. Plugging the above into~\eqref{TD-SI2-full-2} and keeping the leading terms, we see that for $p=0$, the action becomes
\be
\begin{aligned}
\label{TD-SI2-3}
S=\int d^4x \sqrt{g}&\left[\frac{\z \s^{-\frac{1}{4}}+\j h^2}{2}R-\frac{1}{128}\, \s^{-\frac{9}{4}}g^{\m\n}\partial_\m \s\partial_\n \s-\frac{1}{2}g^{\m\n}\partial_\m h \partial_\n h\right. \\
&\left.+\frac{\d}{8}\, \s^{-1}h g^{\m\n}\partial_\m \s \partial_\n h-\frac{\lambda}{4} h^4+\kappa\,\s^{-\frac{1}{4}}h^2-c\, \s^{-\frac{1}{2}}\right] \ ,
\end{aligned}
\ee
with
\be
\label{const-defs}
\xi=\frac{\z f_1}{2} \ ,~~~\k=-c_2\a\ ,~~~\l=4c_2 \b \ ,~~~c=c_2-c_3 \ .
\ee
Making use of~\eqref{sigm-chi-1}, we can express the above in a more familiar form: 
\be
\begin{aligned}
\label{TD-SI2-4}
S=\int d^4x \sqrt{g}&\left[\frac{\z\x^2+\j h^2}{2}R-\frac{1}{2}g^{\m\n}\partial_\m \x\partial_\n \x-\frac{1}{2}g^{\m\n}\partial_\m h \partial_\n h\right. \\
&\left.-\d\, \x^{-1}h g^{\m\n}\partial_\m \x \partial_\n h-\frac{\lambda}{4} h^4+\kappa\,\x^2h^2-c\,\x^4\right] \ .
\end{aligned}
\ee

Notice that once we identify the scalar field $h$ with the Higgs boson (in the unitary gauge), then for $\d=0,$ the above bears resemblance to the phenomenologically viable Higgs-dilaton cosmological model that was constructed in~\cite{Shaposhnikov:2008xb,Shaposhnikov:2008xi} and studied in great detail in~\cite{GarciaBellido:2011de,GarciaBellido:2012zu,Bezrukov:2012hx,Rubio:2014wta}. There are, however, certain differences which should be pointed out. First of all,  in the present context, we need not introduce the field $\chi$ ad hoc, since this degree of freedom is already present in the gravitational sector. Moreover, as we mentioned before, a symmetry-breaking potential is absent. This means that contrary to what happens in theories for which $p\neq 0$, the scale symmetry of the system remained intact when it was cast into a form invariant under the full group of diffeomorphisms. Finally, it is interesting to note that the way  this theory is derived here is much simpler as compared to the conventional SITDiff, where complicated theory-defining functions have to be chosen~\cite{Blas:2011ac}.

Once we have identified $h$ with the Higgs field, we have to make sure that the theory has satisfactory particle physics as well as cosmological phenomenology, which puts constraints on the various parameters that appear in the action~\eqref{TD-SI2-4}. To start with, we observe that we have to set $\lambda\sim \mathcal O(1)$ in order for the model to be compatible with the SM predictions. Also, if $h$ is responsible for the inflationary expansion in the early Universe, then the nonminimal coupling has to satisfy $\xi\approx 47000\sqrt{\lambda}$, such that the amplitude of the primordial fluctuations agree with the observations~\cite{Bezrukov:2007ep}. 

Moreover, since $\kappa$ accounts for the difference between the Higgs boson mass and the Planck mass, it should be fixed at order $\mathcal O(10^{-30})$. In addition, we have to impose $c\sim\mathcal O(10^{-120})$ to reproduce the hierarchy between the value of the cosmological constant and the Planck scale. In the next section, we will present a conjecture about why these two parameters might be zero at the classical level.

\section{Regularity?}
\label{sec:Regul}

The fact that the Higgs boson mass and the cosmological constant terms are much smaller with respect to the Planck scale, might be an indication that at the  level of fundamental action both of them are zero. It is reasonable to wonder whether it exists some underlying principle or mechanism that forbids the presence of these terms in the action. 

Inspection of~\eqref{TD-SI2-3} reveals that due to the peculiar way the dilaton appears, all terms in the action that involve this field seem to be ill defined when $\sigma\rightarrow 0$, arguably related to the high energy limit.  As we will demonstrate in this section,  this is not the case if the theory is expressed in terms of  variables conjugate to space and time derivatives of the fields. These momentum densities were first introduced by Schwinger~\cite{Schwinger:1951xk,*Schwinger:1953tb} (see also~\cite{Heinzl:2000ht}) and should be thought of as the covariant counterparts of canonical momenta.  
For a theory described by a Lagrangian $\mathscr L[\f_i, \partial_\m \f_i]$ which depends on a set of fields $\phi_i$ and their derivatives $\partial_\m \f_i$, these quantities are defined as 
\be
\label{momenta-1}
\p^{\ \m}_i\equiv\frac{\d\mathscr L}{\d \partial_\m\f_i} \ .
\ee
Let us focus now on~\eqref{TD-SI2-3} and set $\d=0$, such that there is no kinetic mixing between the Higgs and the dilaton. This is purely for convenience, since the results will not be qualitatively different from the case where the mixing term is present, whereas the manipulations simplify considerably.  For our purposes, it is necessary to cast the action in such a way that it only contains first derivatives of the metric. A straightforward calculation along the lines of the one in~\cite{landau_fields} for the Einstein-Hilbert action, gives us
\be
\label{TD-SI2-5}
S=\int d^4x \sqrt{g}\,\mathscr L \ ,
\ee
where the Lagrangian $\mathscr L$ is
\be
\begin{aligned}
\label{lagr-1}
\mathscr L&=\frac{\z \s^{-\frac{1}{4}}+\j h^2}{2}T^{\a\b\g\k\l\m}\G_{\a\b\g}\G_{\k\l\m}+\left(\j h \partial_\n h-\frac{\z \s^{-\frac{5}{4}}}{8}\partial_\n \s\right) S^{\k\l\m\n}\G_{\k\l\m}\\
&-\frac{1}{128}\, \s^{-\frac{9}{4}}g^{\m\n}\partial_\m \s\partial_\n \s-\frac{1}{2}g^{\m\n}\partial_\m h \partial_\n h-\frac{\l}{4} h^4+\kappa\,\s^{-\frac{1}{4}}h^2-c\, \s^{-\frac{1}{2}} \ .
\end{aligned}
\ee
Here 
\be
\label{chr-down}
\G_{\l\m\n}=\frac{1}{2}\left(\partial_\n g_{\m\l}+\partial_\m g_{\l\n}-\partial_\l g_{\m\n}\right) \ ,
\ee
and we introduce the tensors
\be
\label{S-T-tensor}
S^{\k\l\m\n}= g^{\k\l}g^{\m\n}-g^{\n\k}g^{\l\m}~~~\text{and}~~~T^{\a\b\g\k\l\m}= g^{\a\l}g^{\b\k}g^{\g\m}-g^{\a\b}g^{\g\k}g^{\l\m} \ .
\ee
Using~\eqref{momenta-1}, we find that  Schwinger's  ``momenta,''
\be
\label{momenta-2}
\p_h^{ \ \n}= \frac{\d\mathscr L}{\d \partial_\n h} \ ,~~~\p_\sigma^{ \ \n}= \frac{\d\mathscr L}{\d \partial_\n \sigma} \ ,~~~\text{and}~~~\r^{\l\m\n}=\frac{\d\mathscr L}{\d \G_{\l\m\n}} \ ,
\ee
are given by 
\be
\label{momenta-3}
\p_h^{\ \n}=\j h S^{\k\l\m\n}\G_{\k\l\m}-\partial^\n h \ , \ \ \ \p_\s^{\ \n}=-\frac{1}{8}\left(\z \s^{-\frac{5}{4}} S^{\k\l\m\n}\G_{\k\l\m}+\frac{1}{8} \s^{-\frac{9}{4}}\partial^\n \s\right) \ ,
\ee
and
\be
\label{momenta-4}
\r^{\l\m\n}=\frac{\z \s^{-\frac{1}{4}}+\j h^2}{2}\left(T^{\a\b\g\l(\m\n)}+T^{\l(\m\n)\a\b\g}\right)\G_{\a\b\g}+S^{\l(\m\n)\a}\left(\j h \partial_\a h-\frac{\z}{8}\s^{-\frac{5}{4}}\partial_\a \s\right) \ ,
\ee
where the parentheses $(\ldots)$ denote symmetrization of the corresponding indices. Using the relations~\eqref{momenta-3}, the above can be rewritten as
\be
\begin{aligned}
\label{momenta-4}
\r^{\l\m\n}&=\frac{\z \s^{-\frac{1}{4}}+\j h^2}{2}\left(T^{\a\b\g\l(\m\n)}+T^{\l(\m\n)\a\b\g}\right)\G_{\a\b\g}\\
&+S^{\l\m\n\d}S^{\a\b\g}_{\ \ \ \ \d}\G_{\a\b\g}\left(\z^2\s^{-\frac{1}{4}}+\j^2h^2\right)-S^{\l\m\n}_{\ \ \ \ \k}\left(\j h \p_h^{\ \k}+8\z \s \p_\s^{\ \k} \right)\ .
\end{aligned}
\ee
In terms of $\r,\ \pi_h, \ \text{and} \ \p_\s$, we find that~\eqref{lagr-1} becomes
\be
\begin{aligned}
\label{momenta-rho-explicit}
\mathscr L&= \left(\frac{1}{\z \s^{-1/4}+\j h^2}\right)\left(\frac{(1+4\z)\z \s^{-1/4}+(1+4\j)\j h^2}{(1+6\z)\z \s^{-1/4}+(1+6\j)\j h^2}\right)\times\\
&\times \left(\frac{1}{4}\r_{\k\l}^{\ \ \l}\r^{\k\m}_{\ \ \m}+\frac{1}{3((1+4\z)\z \s^{-1/4}+(1+4\j)\j h^2)}\r_\k^{\ \k\l}\r^{\m}_{\ \m\l}-\r_\k^{\ \k\l}\r_{\l\m}^{\ \ \m}\right)\\
&+\left(\frac{1}{\z \s^{-1/4}+\j h^2}\right)\left(\r_{\l\m\n}\r^{\m\n\l}-\frac{1}{2}\r_{\l\m\n}\r^{\l\m\n}\right)\\
&+\frac{2\j h}{((1+6\z)\z \s^{-1/4}+(1+6\j)\j h^2)}\left(\r_\k^{\ \k\l}-\frac{1}{2}\r^{\l\k}_{\ \ \k}\right)\p_{h\,\l}\\
&-\frac{16\z \s}{((1+6\z)\z \s^{-1/4}+(1+6\j)\j h^2)}\left(\r_\k^{\ \k\l}-\frac{1}{2}\r^{\l\k}_{\ \ \k}\right)\p_{\s\,\l}\\
&-\frac{48\z\j \s h}{((1+6\z)\z \s^{-1/4}+(1+6\j)\j h^2)}\p_h^{\, \m}\p_{\s\,\m}\\
&-\frac{1}{2}\left(\frac{(1+6\z)\z \s^{-1/4}+\j h^2}{(1+6\z)\z \s^{-1/4}+(1+6\j)\j h^2}\right)\p_{h\,\m}\p_h^{\,\m} \\
&-32\s^{9/4}\left(\frac{\z \s^{-1/4}+(1+6\j)\j h^2}{(1+6\z)\z \s^{-1/4}+(1+6\j)\j h^2}\right)\p_{\s\,\m}\p_\s^{\,\m}\\ 
&-\frac{\l}{4} h^4+\kappa\,\s^{-\frac{1}{4}}h^2-c\, \s^{-\frac{1}{2}} \ .
\end{aligned}
\ee 
It is convenient to introduce at this point
\be
\begin{aligned}
\label{momenta-5}
P_{\l\m\n}&\equiv\r_{\l\m\n}-\frac{1}{2}\left(\frac{(1+4\z)\z  \s^{-\frac{1}{4}}+(1+4\j)\j h^2}{(1+6\z)\z  \s^{-\frac{1}{4}}+(1+6\j)\j h^2}\right)g_{\m\n}\r_{\l\k}^{\ \ \k}\\
&-\frac{1}{3}\left(g_{\l\m}\r^\k_{\ \k\n}+g_{\l\n}\r^\k_{\ \k\m}-\frac{\z  \s^{-\frac{1}{4}}+\j h^2}{(1+6\z)\z  \s^{-\frac{1}{4}}+(1+6\j)\j h^2}g_{\m\nu}\r^\k_{\ \k\l}\right)\\
&+\frac{(\z  \s^{-\frac{1}{4}}+\j h^2)}{(1+6\z)\z  \s^{-\frac{1}{4}}+(1+6\j)\j h^2}g_{\m\n}\left(\j h\p_{h\,\l}-8\z \s \p_{\s\,\l}\right) \ ,
\end{aligned}
\ee
such that~\eqref{momenta-rho-explicit}, simplifies considerably and reads
\be
\begin{aligned}
\label{theory-5}
\mathscr L&=\frac{1}{(\j h^2+\z \s^{-\frac{1}{4}})}\left(P_{\l\m\n}P^{\m\n\l}-\frac{1}{2}P_{\l\m\n}P^{\l\m\n}\right)\\
&-\frac{(1+4\j)\j h^2+(1+4\z)\z \s^{-\frac{1}{4}}}{(\j h^2+\z \s^{-\frac{1}{4}})^2}\left(P_\k^{\ \k\l}P_{\l\m}^{\ \ \m}-\frac{1}{2}P_{\k\l}^{\ \ \l}P^{\k\m}_{\ \ \m}\right)\\
&+\frac{2(\j^2 h^2+\z^2 \s^{-\frac{1}{4}})}{(\j h^2+\z \s^{-\frac{1}{4}})^2} P_\k^{\ \k\l}P^\m_{\ \m\l}-\frac{1}{2}\p_{h\, \m}\p_h^{\ \m} -32 \s^{\frac{9}{4}} \p_{\s\, \m}\p_\s^{\ \m}\\
&-\frac{\l}{4} h^4+\kappa\,\s^{-\frac{1}{4}}h^2-c\, \s^{-\frac{1}{2}} \ .
\end{aligned}
\ee
Observe that in the limit where the four-momenta $P$ (or equivalently $\rho$) are kept fixed while $\s$ tends to zero, i.e. for
\be
\begin{aligned}
\s &\rightarrow 0 \ , \\
\p_h \,,\p_\s&\,,\,P~\text{or}~\rho\rightarrow\text{fixed} \ ,
\end{aligned}
\ee
the only terms that blow up are the Higgs mass and the cosmological constant. Therefore, it is tempting to speculate that both these terms should not be included in the action in the first place if we want the theory to remain regular at the UV limit. It is interesting to note that the pathological behavior of the cosmological constant persists for arbitrary metric dimensions, the reason being that it is always proportional to $\s^{-\frac{1}{2}}$. On the other hand, if $p$ is not chosen to be equal to zero, the Higgs mass term, as well as the term proportional to $\s^{\frac{1}{2(p-1)}}$ (which does not feed into the cosmological constant unless $p=0$), are singular only if $p<1$.

Even though we do not have an answer to what is the origin of this selection rule, it could be a manifestation of some yet unknown mechanism at very high energies. Notice that if a scale-invariant regularization scheme is used (see for example~\cite{Shaposhnikov:2008xi}), then these terms cannot be generated at any order in perturbation theory. It may well be the case that they emerge from nonperturbative physics, something that can explain their smallness.

\section{Conclusions}
\label{sec:conclus}

The purpose of this paper was to investigate a previously unexplored region of the parameter space of theories with dilatational symmetry whose gravitational sector is constructed by requiring invariance under the group of transverse diffeomorphisms. Due to the invariance under this restricted group of coordinate transformations, the determinant of the metric becomes a dynamical degree of freedom which can be thought of as a dilaton. 

We argued that the most appropriate and natural option for the description of arbitrary coordinate systems is for the metric to have dimensionality of area. We demonstrated that the particular setup is distinct from the ordinary theories in a number of aspects. The form of the pure gravitational action is completely fixed and moreover, once diffeomorphism invariance is restored via the St\"uckelberg mechanism, the scale symmetry remains intact. As a result, there is no runaway potential for the dilaton. 

Next, we investigated the form of the action of a model that on top of the dilaton contains an extra scalar field which we identified with the Standard Model Higgs boson. Based on the way the dilaton appears and interacts with the Higgs field, we observed that the Higgs mass and cosmological constant are the only singular terms in the specific limit (fixing the proper variables which we define) involving a metric determinant going to zero.  An appealing hypothesis is  that these terms should not be included in the fundamental theory, but rather their low-energy presence should result from nonperturbative effects through some yet unknown mechanism.

It would be interesting to understand how these considerations can be applied to theories without Lorentz invariance, such as, for example, in Ho\v{r}ava-Lifshitz gravity (see for example~\cite{Horava:2009uw,*Blas:2009qj}), a version of which has recently been proven to be renormalizable~\cite{Barvinsky:2015kil}.

\section*{Acknowledgements}

The authors thank Diego Blas for valuable comments. The work of G.K.K. and M.S. is supported by the Swiss National Science Foundation.

\appendices

\section{Dimensional Analysis}
\label{sec:dim-an}

When the metric $g_{\m\n}$ is dimensionful, the operation of lowering and raising indices has to be done with some care, since covariant and contravariant tensors carry different dimensions. For example, the inverse metric $g^{\m\n}$ has dimensions of $\left[\text{GeV}\right]^{2q}$. Moreover, for the metric determinant $g\equiv-\det(g_{\m\n})>0$, we obtain
\be
\label{metdet-dim}
\left[g\right]=\left[\text{GeV}\right]^{-8q} \ ,
\ee
whereas from~\eqref{assign-dims}, it follows that 
\be
\label{der-dim}
\left[\partial_\m\right]=\left[\text{GeV}\right]^{p} \ .
\ee
We are now in position to determine the dimensionality of various geometrical quantities. First of all, for the Christoffel symbols which are defined as
\be
\label{chris-def}
\G^\l_{\m\n}=\frac{1}{2}g^{\k\l}\left(\partial_\n g_{\m\k}+\partial_\m g_{\k\n}-\partial_\k g_{\m\n}\right) \ ,
\ee
we obtain
\be
\label{chris-dim}
\left[\G^\lambda_{\m\n}\right]=\left[\text{GeV}\right]^p \ ,
\ee
in accordance with~\eqref{der-dim}.
Consequently, for the curvatures
\be
\label{riem-def}
R^\k_{\l\m\n}=\partial_\m\G^\k_{\l\n}-\partial_\n\G^\k_{\l\m}+\G^\r_{\l\n}\G^\k_{\r\m}-\G^\r_{\l\m}\G^\k_{\r\n} \ ,\ R_{\m\n}= R^\kappa_{\m\kappa\n} \ ,\ R= g^{\m\n}R_{\m\n} \ ,
\ee
we see that
\be
\label{riem-dim}
\left[R^\kappa_{\ \lambda\m\n}\right]=\left[\text{GeV}\right]^{2p} \ ,~~~\left[R_{\m\n}\right]=\left[\text{GeV}\right]^{2p} \ ,~~~\left[R\right]=\left[\text{GeV}\right]^{2\left(p+q\right)} =\left[\text{GeV}\right]^2\ .
\ee

\bibliographystyle{utphys}
\bibliography{SI_TDiff}{}

\end{document}